\pdfoutput=1 
\documentclass{JINST}
                     \let\ifpdf\relax
\usepackage{color,amsmath}
\usepackage[normalem]{ulem}
\usepackage{lineno}
\usepackage{url}

\title{\boldmath A 20-Liter Test Stand with Gas Purification for Liquid Argon Research}

\author{Yichen Li$^a$\thanks{Corresponding author: yichen@bnl.gov
    (Yichen Li)},
    Craig Thorn$^a$,
    Wei Tang$^a$,
    Jyoti Joshi$^a$,
    Xin Qian$^a$,
    Milind Diwan$^a$,
    Steve Kettell$^a$,
    William Morse$^a$,
    Triveni Rao$^b$,
    James Stewart$^a$,
    Thomas Tsang$^b$, and
    Lige Zhang$^a$\thanks{Also from Drexel University}\\
        \llap{$^a$}Physics Department, Brookhaven National Laboratory,
    Upton, NY 11973, USA\\
        \llap{$^a$}Instrumentation Division, Brookhaven National
        Laboratory, Upton, NY 11973, USA
    }

\abstract{We describe the design of a 20-liter test stand constructed to
study fundamental properties of liquid argon (LAr). This system utilizes
a simple, cost-effective gas
argon (GAr) purification to achieve high purity, which is necessary
to study electron transport properties in LAr. An electron drift stack with up
to 25 cm length is constructed to study electron drift, diffusion, and attachment at
various electric fields. A gold photocathode and a pulsed laser are used as a bright
electron source. The operational performance of this system is reported.}


\begin{document}
\maketitle
\flushbottom
\section{Introduction}

Research and development (R\&D) of the liquid argon time projection chamber (LArTPC) is 
at the technical frontier of neutrino physics. As the chosen technology of the 
short-baseline neutrino (SBN) program at Fermi National Accelerator Laboratory
(FNAL)~\cite{Antonello:2015lea} 
and the Deep Underground Neutrino Experiment (DUNE) at the Long-Baseline Neutrino
Facility (LBNF)~\cite{Adams:2013qkq}, LArTPCs provide a key technology to search for light 
sterile neutrino(s)~\cite{Abazajian:2012ys}, to search for CP violation
in the neutrino sector~\cite{Adams:2013qkq}, and to determine the
neutrino mass hierarchy~\cite{Qian:2015waa}. Therefore, the fundamental
properties of LAr are of particular interest for experimentalists. 

Based on the development of the LAr ionization chamber~\cite{willis74}
and the gas-filled time projection chamber~\cite{Nygren:1976fe}, LArTPCs
were introduced to neutrino physics in the period 1976 to 1978 by H. H.
Chen~\cite{Chen:1976pp} and by C. Rubbia~\cite{rubbia77} as a
fine-grained tracking calorimeter. In LArTPCs, energetic charged
particles produced in neutrino interactions ionize argon atoms as they
traverse the LAr medium. These ionization electrons then travel along an
externally applied electric field ($\sim$ 500 V/cm) at a constant
velocity towards position-sensitive detectors at the anode plane. The
time difference between the creation of the ionization electrons  and
their arrival time, the arrival position, and the charge in the electron
swarm carry important three-dimensional position and energy information
of the initial energetic charged particles. The ionization electrons
typically take a few  milliseconds to drift a few meters in large
neutrino detectors. During this drift, each ionization electron
interacts continuously with atoms and molecules along its path. If
electronegative impurities are present in the LAr, drifting electrons
will be captured to form negative ions, which drift much more slowly
than electrons, resulting in a decrease in the fast component of the
charge detected at the anode.  This reduces the signal-to-noise ratio
and makes the collected charge dependent on drift distance in addition
to ionization. In general, the free electron lifetime is used to
characterize this attachment process, and a free electron lifetime
longer than the drift time is typically required to avoid losing charge
information.  Therefore, high purity LAr with less than
part-per-billion (ppb) electronegative impurities is essential. This
level of impurity is
usually several orders of magnitude lower than that of commercial LAr.
Thus, a dedicated argon purification system is essential to study
electron transport properties of LAr.  
 
Most existing LAr test facilities, such as LAPD~\cite{Adamowski:2014daa}
and MTS~\cite{Andrews:2009zza} at FNAL, and
ARGONTUBE~\cite{Ereditato:2013xaa} at Bern, utilize LAr purification to
remove impurities (O$_2$ and H$_2$O) directly from the liquid  by
recirculating it through a purifier. This technique has the advantage
that a large mass of argon can be processed in a short time. However,
the implementation of LAr purification usually requires a
cryogenic pump, which can be expensive to purchase and
maintain and requires significant LAr plumbing.  In this paper, we
report the design and operational performance of a simple,
cost-effective 20-liter liquid test stand constructed at Brookhaven
National Laboratory (BNL), which relies solely 
on gas purification to achieve high purity LAr. This test stand is
constructed specifically to measure electron transport properties in
LAr. It is an upgrade of a liquid argon test stand reported in
Ref.~\cite{Li:2015rqa}, providing better thermal stability and an
improved drift stack apparatus.

This paper is organized as follows. In Sec.~\ref{sec:description}, we
describe various components of our test stand including the cryogenic
system, the laser and the photocathode, the electron drift stack, and
the data acquisition system. In Sec.~\ref{sec:performance}, we report
the performance of our test stand including the thermal stability,
    purity, high voltage capability, and electron source
    characteristics.  In Sec.~\ref{sec:future}, we will discuss future measurements and provide a
    summary.

\section{System Description}\label{sec:description}
\subsection{Cryogenic System}\label{sec:cryo_des}

\begin{figure}[htbp] 
\begin{minipage}[b]{0.6\linewidth}
\centering
\includegraphics[width=1.0\textwidth,angle=0]{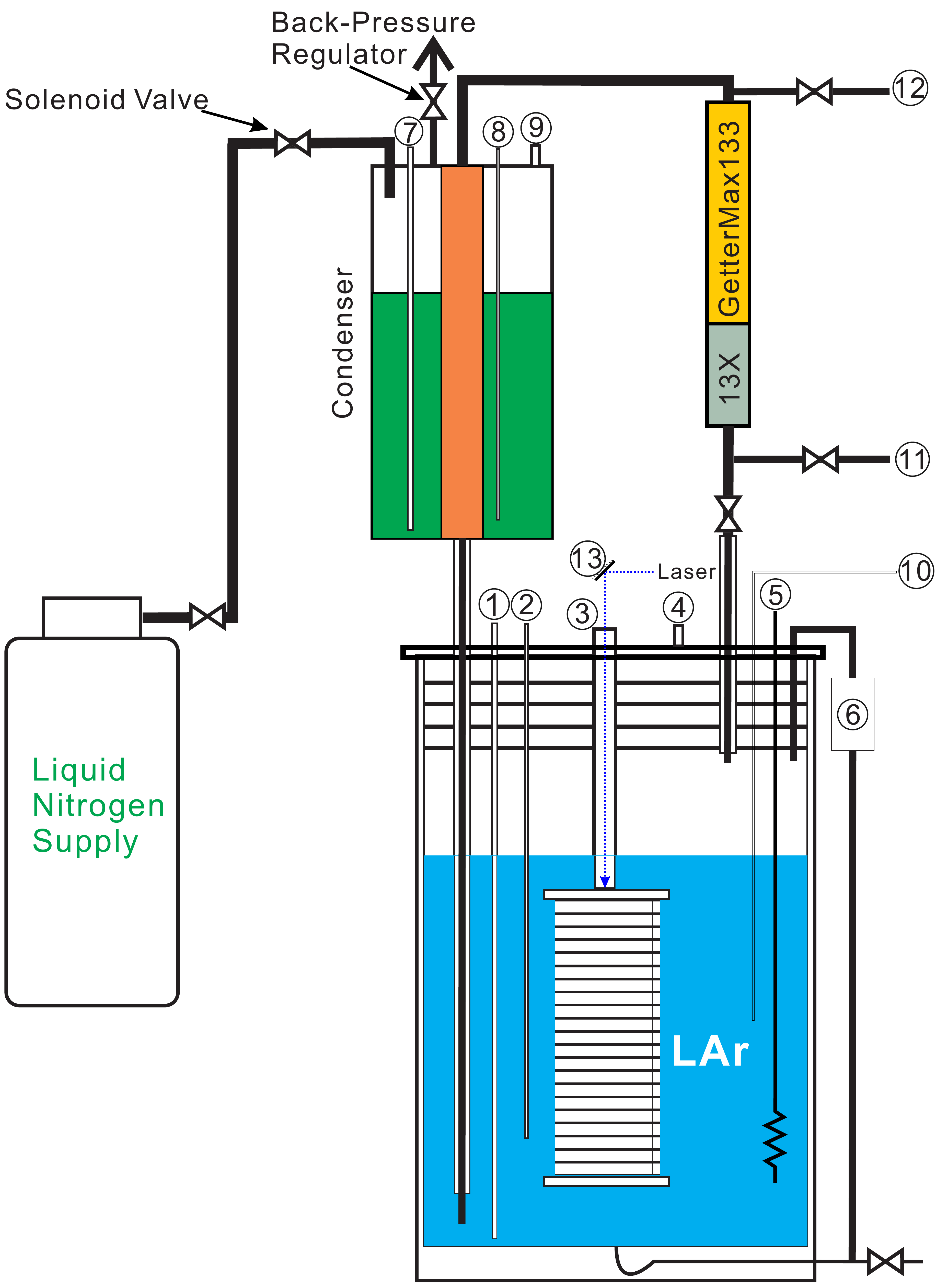}
\end{minipage}
\begin{minipage}[b]{0.4\linewidth}
\begin{tabular}[b]{l}
1. capacitance level gauge\\
2. resistance temperature detector\\(RTD) probe\\
3. optical tube for laser beam entrance\\
4. pressure transducer\\
5. heater\\
6. pressure differential level gauge\\
7. capacitance level gauge\\
8. RTD probe\\
9. pressure transducer\\
10. sampling tube to gas analyzer\\
11. activation gas inlet\\
12. activation gas outlet\\
13. dielectric mirror\\
\end{tabular}
\end{minipage}
\caption{A schematic diagram of the apparatus. The components are
    labeled
with numerical indices. 
HV and signal feedthroughs are omitted for the display and are discussed in the text.}
\label{fig:schematic}
\end{figure}

The cryogenic system is composed of a 20-liter LAr dewar, 
a  LAr condenser that is cooled with pressurized liquid nitrogen at
$\sim$ 87 K, an argon purifier containing an oxygen absorber and a water
adsorber, and the associated plumbing that connects these components. A
schematic diagram of the apparatus is shown in Fig.~\ref{fig:schematic}.

The main 20-liter LAr dewar is manufactured by CryoFab (part \# CF9524)
with multi-layer insulation (MLI)~\cite{cryofab}.  It has been pressure
tested to 30 psig but the maximum allowable working pressure (MAWP) is
limited to 8 psig by a relief valve to meet BNL safety requirements.
The cylindrical dewar has internal dimensions of 24 inches in depth by
9.46 inches in diameter with an ellipsoidal bottom, for a total (empty)
internal volume of 27.9 liters.  A 12-inch diameter conflat flange,
containing 15 feedthroughs, is used to seal the dewar at the top. These
feedthroughs provide connections for the gas outlet and purified liquid
inlet as well as all the test instruments including an evacuated optical
tube for laser beam entrance, high and low voltage supply cables, a
level gauge, temperature and pressure sensors, and signal cables. Four
thin stainless steel baffle plates spaced at 1 inch intervals are installed below
the top flange to provide thermal insulation between the LAr and the top flange.
This set of baffles significantly reduces the heat transfer due to
radiation and GAr convective flow. The effectiveness of the insulation
is confirmed by temperature measurements of the upper surface of the top
flange with LAr filled just below the bottom plate. The surface
temperature remains at about 20 $^{\circ}\rm C$, preventing water
condensation on the top of the flange. The LAr volume below the lowest
baffle plate is 22.1 liters. A tube passing through the insulating
vacuum space is installed at the bottom of the dewar as a port to fill
and drain LAr and a connection to the bottom of the LAr to measure the
head pressure.

In a LArTPC, LAr with high purity is essential in order to observe signals
from drifting ionization electrons. Electronegative molecules, such as oxygen and
water, attach drifting electrons. The resulting negative ions drift very
slowly ( $\sim$ 10$^5$ times slower than electrons) and the slow signals they
induce at the readout are outside the bandwidth of the detection
electronics. In large systems, impurities have to be reduced to $\sim$
0.1 ppb of oxygen equivalent to keep this attenuation of the signal to
an acceptable level. In small systems, this requirement  may be relaxed due to the shorter
    free electron drift times typically involved.  In our system, the purifier is
a 304 stainless steel tube, 13 inches long with a diameter of 2 inches,
  closed at the top by a conflat flange.  The bottom is connected to a
  vacuum-jacketed tube that penetrates the top flange of the main dewar
  and conducts GAr into the purifier from the space above the LAr and
  below the baffle plates. It is filled with 1 kg of 13X molecular sieve
  8-14 mesh beads and 3 kg of GetterMax-133 copper catalyst  3 $\times$
  3 mm tablets \cite{gettermax} for removing water and oxygen, respectively. The
  molecular sieve is filled at the bottom of the cylinder, and the
  GetterMax-133 is filled on the top of it.  Water is removed by the
  selective molecular adsorption in the pores of the molecular
  sieve~\cite{mavrok}. Oxygen is removed by the oxidative reaction with
  the copper. Both processes reduce the purification capacity as the
  active materials become saturated with the impurities.  A regeneration
  procedure is required to restore the capacity of the materials before
  initial usage and after several fillings of the system with commercial
  LAr.  This procedure is described in detail in Sec.~\ref{sec:cryo_op}.
  Appropriate valving of the purifier allows the regeneration to occur
  in situ.

\begin{figure}[htbp] 
\centering
\includegraphics[width=0.6\textwidth, angle=0]{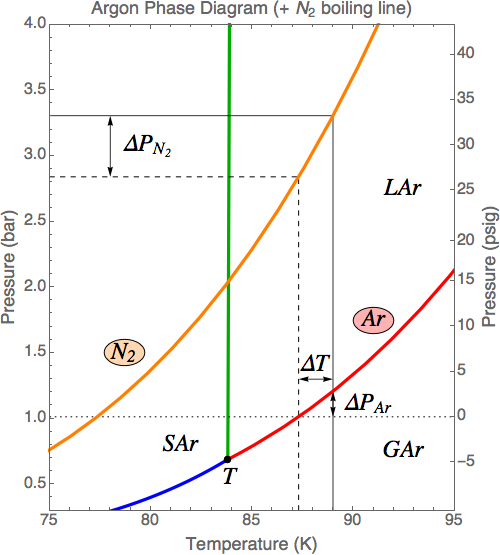}
\caption{Phase diagram of argon with the boiling line of N$_2$ superposed. 
LN$_2$ is the only source
of cooling in this system. The dashed line marks the normal boiling point of LAr at 87.3 K and the corresponding N$_2$ pressure. The 
solid line indicates the normal operating temperature of 89 K, at $\Delta$T above the normal boiling point.
$\Delta$P$_\text{Ar}$ and $\Delta$P$_{\text{N}_2}$ indicate the
pressure differences due to the temperature difference $\Delta$T. 
The ratio of LN$_2$ to LAr pressures ranges from 2.00 to 2.66 over the
operating temperature range of 87.3 K to 91.2 K.}
\label{fig:heat_ex}
\end{figure}

The argon condenser is constructed of 304 stainless steel.  An outer
cylinder 6 inches in diameter and 18.5 inches in length contains the
pressurized LN$_2$, which is the source of refrigeration.  The filling
of the LN$_2$ is controlled by a cryogenic solenoid valve (Gems
        D2063-LN2) with feedback of the LN$_2$ level as measured by a
capacitance level gauge. During operations, GAr evaporated from the LAr
is condensed in an inner coaxial tube 2 inches in diameter, which is
packed with coarse copper wool to improve the heat transfer between the
GAr and the LN$_2$ outside the inner tube. The entire condenser is
insulated with 6 inches of Cryo-Lite cryogenic insulation~\cite{cryolite}.
Fig.~\ref{fig:heat_ex} shows the phase diagram of argon with the boiling
line of N$_2$ superposed. The LN$_2$ temperature is controlled by the
pressure inside the condenser, which is maintained by a high precision,
adjustable back pressure regulator (Equilibar EB1HF2-SS)~\cite{bpr}. The
adjustment of the regulator sets the LN$_2$ temperature, which in turn
determines the LAr temperature and thus its pressure, as is indicated in
Fig.~\ref{fig:heat_ex}.  The condensed argon flows back to the bottom of
the LAr dewar through a vacuum-jacketed tube penetrating the top flange.
In order to increase the flow rate of the argon through the purifier, a
heater with a resistance of 6 $\Omega$ and 150 W maximum power
dissipation is installed to one side near the bottom of the dewar.

The piping of the system is composed mainly of the LN$_2$ filling tube,
the LAr circulation loop, filling tube, and sampling tubes as shown in
the schematic diagram in Fig.~\ref{fig:schematic}. The LN$_2$ filling
tube introduces LN$_2$ to the LAr condenser through a flexible hose
covered with 2-inch thick Cryo-Lite insulation. In the LAr circulation
loop, LAr is vaporized by heat influx into the liquid through the
imperfect insulation of the dewar walls. The gas produced flows through
the vacuum-jacketed tube with its inlet below the baffle plates. After
passing through the purifier, the gas exchanges heat with the copper
wool cooled by the LN$_2$ and is condensed back into pure liquid. The
liquid then flows to the bottom of the condenser and back into the
bottom of the main dewar by gravity. The heat of the GAr condensation is
removed by the vaporization of LN$_2$. Controlling the pressure of the
LN$_2$, controls both the temperature and the pressure of LAr, following
the boiling curve (red line in Fig.~\ref{fig:heat_ex}).  The LAr
circulation loop components that are not vacuum-jacketed are insulated
by 6-inch thick of Cryo-Lite. LAr is introduced into the dewar through
the bottom filling tube from the supply dewar. Argon sampling lines are
used to conduct gas samples into three gas analyzers (one each for
H$_2$O, N$_2$, and O$_2$) from three locations (downstream and upstream
of the purifier, and from a 1/4-inch tube terminated in the LAr with a
1/32-inch orifice).

Crucial system operating parameters including temperatures, pressures, and 
levels of LAr and LN$_2$ are continuously monitored throughout the entire 
operation. The temperatures for LAr and LN$_2$ are measured by PT-100
RTD probes (Omega PR-10-2) with $\pm$~0.01K precision~\cite{omegartd}.
These probes are calibrated in LN$_2$ at the factory.  A thermocouple is
installed on the surface of the purifier to monitor the temperature
during operation and activation. Another RTD sensor is installed on the
top of the resistors in the heater (about 4 inches above the  level of
the RTD probe measuring LAr temperature) to provide another LAr
temperature reading during operation. The pressures of LAr and LN$_2$
are measured by Omega PX-209 transducers with a specified $\pm$ {0.25\%}
    accuracy~\cite{omega}.  The level of LAr and LN$_2$ are measured by
    coaxial capacitance level gauges (Sycon SLL-N2). In addition, a
    differential pressure transducer (GP:50 216)~\cite{gp50} with
    $\pm$~0.01 cm specified precision is installed to monitor the LAr
    level by measuring the pressure difference between the bottom and
    top of the dewar.  The water concentration is measured by a Servomex
    DF-750e gas analyzer with a lowest detection level of 0.2 ppb. The
    oxygen and nitrogen concentrations are monitored by a Servomex
    DF-560e gas analyzer, with a lowest detection level of 0.2 ppb, and
    a Servomex K2001, with a lowest detection level of 10 ppb,
    respectively~\cite{servomex}.

\subsection{Electron Source: Photocathode and Laser System}
\begin{figure}[htbp]
\centering
\includegraphics[width=0.6\textwidth, angle=0]{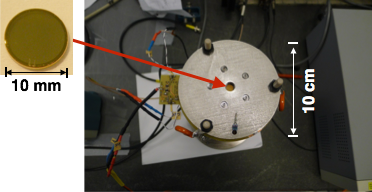}
\caption{Photocathode, mounted in a stainless steel holder, at the top of the drift stack.}
\label{fig:cathode}
\end{figure}

\begin{figure}[htbp]
\centering
\includegraphics[width=0.6\textwidth, angle=0]{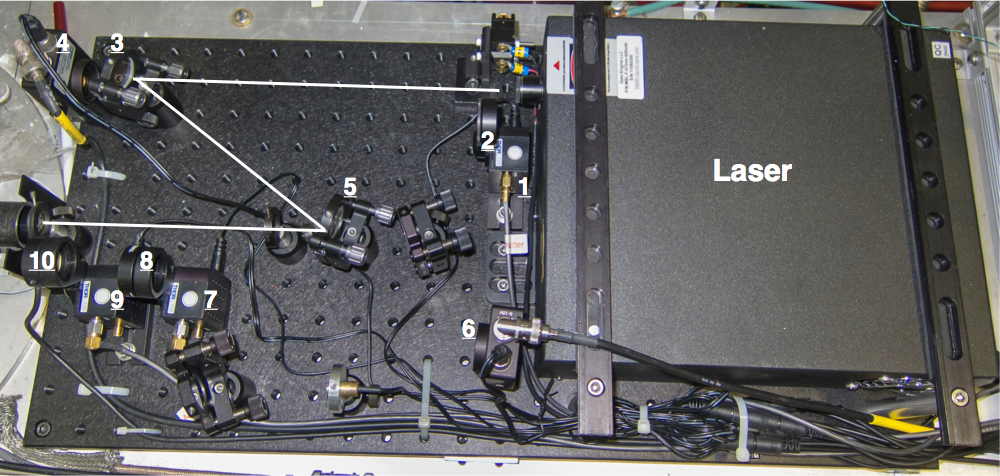}
\caption{The configuration of laser
    components is shown. The light path of the UV laser is drawn with
        straight lines. The essential components are labeled with
        numeric indices in their appearance sequence on the light
        path. 1), 7) and 9) are the mechanical flippers with optical part
        holders controlled by TTL gate signals. 2), 8) and 10) are
        the beam block, attenuator, and power meter mounted on
        individual mechanical flippers, respectively. 3) and 5) are
        dielectric
        mirrors which reflect the laser beam to form its light path. 4)
        and 6) are fast photodiodes behind the dielectric mirrors
        which detect the leakage photons for the trigger.}
\label{fig:laser}
\end{figure}

A photocathode immersed in the LAr, illuminated by a laser system, is
used as a bright electron source. The laser-driven source typically has
a small spatial dimension, on the order of $10^2$ $\mu$m, and a pulse
width on the order of 1 ns. The semi-transparent photocathode
($\sim$~50\% UV transmission) is a 22-nm thick Au film  evaporated onto
a 1 mm thick, 10 mm diameter sapphire disk mounted in a stainless steel
holder at the top of the drift stack as shown in Fig.~\ref{fig:cathode}.
Photoelectrons are generated by irradiating the photocathode with a
frequency quadrupled, 266 nm wavelength (4.66 eV) Nd:YAG laser (CNI
MPL-F-266). As shown in Fig.~\ref{fig:laser}, the laser is installed on
an optical breadboard above the top flange of the dewar. The light path
is formed with two dielectric mirrors.  Along the light path, three
mechanical flippers (Thorlab MMF001) are mounted in sequence to move a
beam block,  beam attenuator, and power meter (Thorlab S120VC) in and
out of the light path. The mechanical flippers are controlled by the
slow control system. The laser pulse is triggered by a signal generated
by the control system.  The laser operates stably at repetition rates
between 1 kHz and 5 kHz. The trigger to the data acquisition system is
initiated by one of the fast silicon photodiodes (Thorlab DET10A)
located behind each dielectric mirror that receives the leakage photons.
Photoelectrons are emitted into the LAr from the surface of the Au
photocathode that is back-illuminated by the laser through an evacuated
optical feedthrough tube sealed with sapphire optical windows.  The
laser beam is deflected into the optical feedthrough tube and steered
onto the photocathode by a dielectric mirror at the top of the dewar as
shown in Fig.~\ref{fig:schematic}. A lens installed between the mirror
and the top sapphire window focuses the laser to a spot typically $\sim$
150 $\mu$m in diameter.  The laser energy density on the photocathode is
typically one order of magnitude lower than the damage threshold,
measured to be about 6 mJ/cm$^2$.

\subsection{Electron Drift Stack}

A drift stack has been constructed for the electron transport properties
measurements, which is an improved version of the drift stack described
previously in Ref.~\cite{Li:2015rqa}. The field uniformity has been
improved by a redesign of the rings and the cathode holder, and  the
charge noise has been reduced by integrating the pre-amplifier into the
anode board and operating it in the LAr.  The structure is similar to
the LAr purity monitor implemented by ICARUS \cite{monitor}. The
photocathode holder assembly is composed of a 2 mm thick stainless steel
plate with 10 mm aperture and 0.2 mm lip to hold the photocathode, which
is clamped by another 1 mm thick stainless plate with 9.0 mm aperture.
The photocathode is installed in the holder assembly with electric
contacts to the holder as shown in Fig.~\ref{fig:cathode}.  Negative
high voltage is applied to the holder assembly. Field shaping rings with
1 mm thickness by 3.5 inch outer diameter and a 2.5 inch inner diameter
are spaced 6 mm apart with three spacers made in Polyetheretherketone
(PEEK) between each stage. A series of 1 G$\Omega$ cryogenic compatible
resistors make electrical contact with each ring and with the
photocathode.  The cathode and anode regions are separated by a grid
made of stainless steel mesh with a wire width of 17 $\mu$m on a 64
$\mu$m pitch. This mesh is sandwiched between two 500 $\mu$m thick G10
plates using Micro-Megas manufacturing technique~\cite{giomataris}. The
grid is installed on top of the anode board to screen the slow rise
signal induced on the anode by the drifting electrons. The anode is a 1
inch diameter copper disk on a 3.5 inch diameter printed circuit board
with the pre-amplifier circuit integrated at the outside edge of the
same board. Drift distances from 6.5 mm to 250 mm in 7 mm steps can be
achieved by adding or removing field shaping rings between the anode
board and the photocathode. 

\begin{figure}[htbp]
\centering
\includegraphics[width=0.7\textwidth, angle=0]{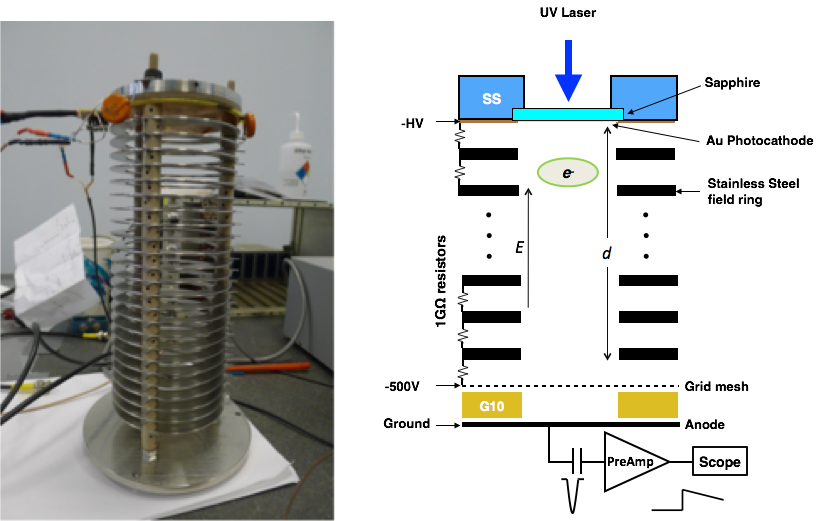}
\caption{The electron drift stack (left) and the principle of electron
    transport properties
    measurement (right) are shown.}
\label{fig:cathode1}
\end{figure}

\subsection{High Voltage Power Supplies}\label{sec:hvs}
Two high voltage power supplies are used in the system.  The potential
for the drift field is supplied by a Glassman model ER30 with an
adjustable voltage range up to -30 kV.  A Bertan model 1170 with a
voltage range  up -5 kV is used to establish the collection field
between the grid mesh and the anode. In order to avoid collecting
drifting electrons on the grid, the collection field must be
significantly higher than the drift field. The collection field is
typically set at 10kV/cm, which ensures grid transmission greater than
80\% for drift fields below 1 kV/cm.  The transmission has been computed
by drifting electrons in an electric fields generated with Maxwell 3D
for the measured grid geometry. The Glassman supply is connected to the
drift stack with a custom HV feedthrough, consisting of a RG-8 cable
cast into an epoxy housing that seals to the top flange. To avoid
breakdown in the GAr, the conductor of the cable inside the dewar is
hermetically sealed in Teflon insulation. The Bertan supply connection
is through a commercial SHV conflat feedthrough mounted on the top
flange and connected to a RG-58 cable. 

\subsection{Slow Control and Data Acquisition System}

The slow control system consists of five functions. They are 1)
monitoring and recording of system parameters, 2) controlling the HV
supplies to establish specified drift and collection fields, 3) filling
and level monitoring of the LN$_2$ condenser, 4) laser and trigger
control, and 5) system malfunction notification via email.  Each
function is implemented by an individual LabVIEW Virtual Instrument (VI)
program. Three DAQ modules produced by National Instruments are utilized:
two NI-6259 modules each with 16 channels of analog input/output at
16-bit precision and one NI-9217 module with 4 channels of RTD input at
24-bit precision. The NI-6259 module digitizes the analog output of
sensors monitoring the key performance parameters of the system
including pressures and levels of LAr and LN$_2$.  It also provides the
analog output signal to drive the laser. The NI-9217 digitizes the
various temperatures measured by RTD sensors. The cryogenic system
monitoring VI continuously monitors and records the key performance
parameters of the system including the pressures, temperatures, and
levels of both LAr and LN$_2$ every 60 seconds.  The HV supply
controller VI programs the voltages of the Glassman and Bertan supplies
and monitors the voltage, current, and trip protection. The automatic
condenser filling VI opens and closes the solenoid valve between the
LN$_2$ supply tank and the condenser with feedback from the level gauge
to maintain the LN$_2$ level in the condenser between specified set
points. The laser driver VI provides the TTL pulse to trigger the laser
and controls the mechanical flippers. The system alarm notification VI
sends an email notification when the liquid nitrogen supply tank is
empty. 

As shown in Fig.~\ref{fig:cathode1}, the charge of electrons arriving at
the anode is converted to voltage by a charge sensitive pre-amplifier
(BNL IO-538) which is mounted on the anode board immersed in LAr. This
pre-amplifier has a sensitivity of  1 mV/fC, and a RMS noise of  1000
electrons at 87 K. The output of the pre-amplifier is amplified and
shaped by an Ortec  572  NIM module.  The voltage signals from the
preamplifier, the shaping amplifier, and the fast silicon photodiode
sampling the laser beam are recorded on an oscilloscope (Tektronix
4034B) triggered by the photodiode. The logic of the DAQ system is
implemented as a LabVIEW VI which controls both the HV power supplies
and the oscilloscope. It reads a table of HV settings and acquires and
records data at each HV setting. A typical data set at 40 electric fields can be
acquired in 15 minutes.  The temperature variation over this period is
less than 0.1 K.

\section{System Operation and Performance}\label{sec:performance}
\subsection{Cryogenic System Operation}\label{sec:cryo_op}
\begin{figure}[htbp] \centering \includegraphics[width=0.9\textwidth,
    angle=0]{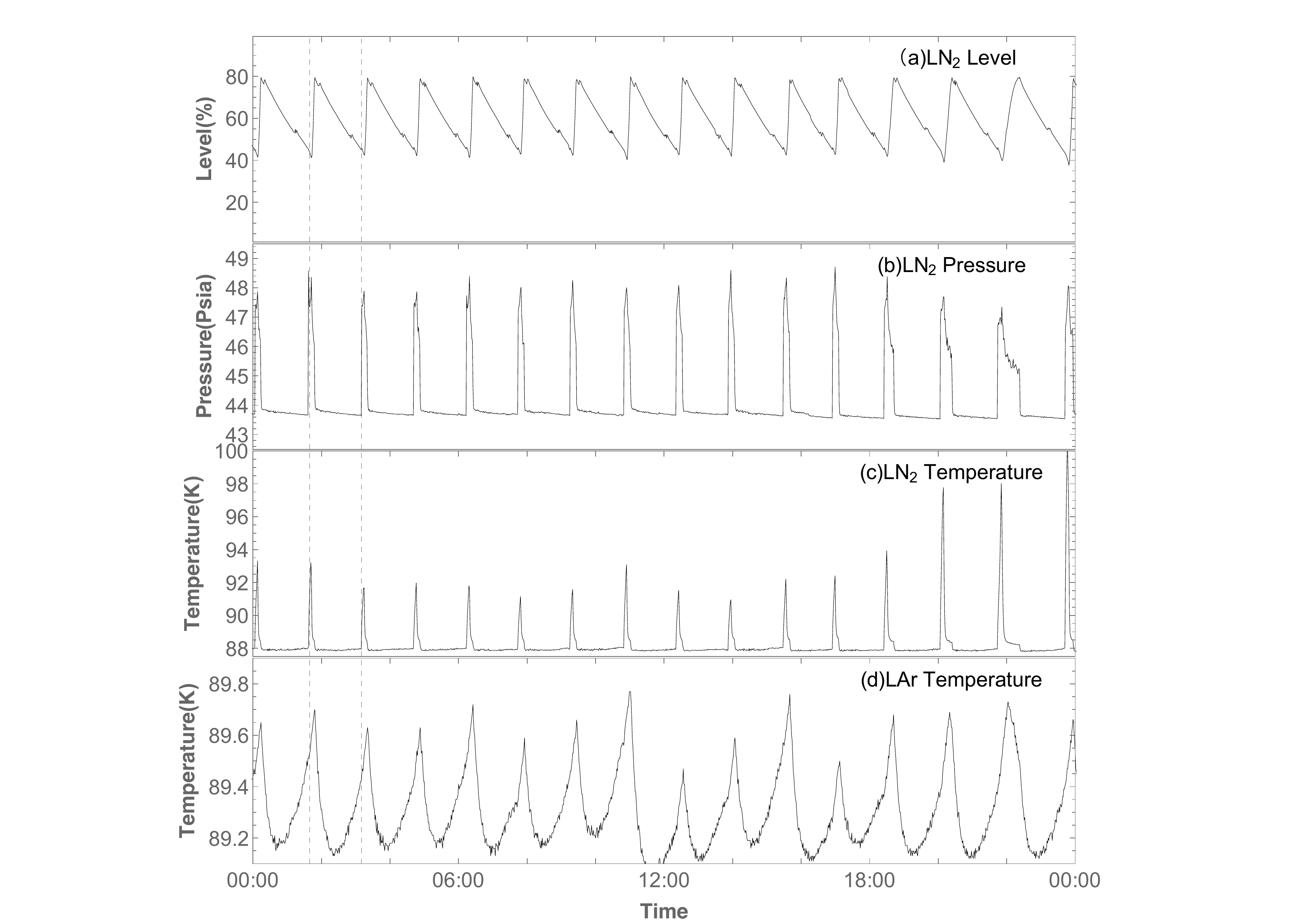} \caption{The cryogenic system
        operating parameters including: (a) liquid nitrogen level in the
        condenser by volume, (b) liquid nitrogen pressure in the
        condenser, (c) liquid nitrogen temperature, and (d) LAr
    temperature are displayed during one day of the operation. The time
range with the dashed line represents one filling cycle. See text for
further discussions.} \label{fig:lnlevel} \end{figure}

The standard operation of the cryogenic systems involves the following steps. 
\begin{itemize}
\item {\bf  Testing for Leaks} \\
Before filling with cryogens, the system is evacuated and tested for
leaks with a vacuum helium leak check. This is necessary since leakage
of air into the system is a source of impurities that lead to a loss of
drifting electrons.  
This test requires no detectable helium leakage into the system with a calibrated 
detector with 1 $\times$ 10$^{-8}$ standard cc/sec. sensitivity.
\item {\bf Pumping and Purging} \\
Before filling with LAr, the system is evacuated to $\sim2\times10^{-5}$ 
Torr and then filled with GAr with 99.999\% purity to about 7 psig at room temperature 
in order to reduce the contamination in the dewar.
This pump-and-purge cycle is repeated 2 - 3 times
to dilute any contamination.
\item {\bf Cooling and Filling} \\
        The condenser is cooled to $<$ 100 K with 
LN$_2$ to allow the condenser to be quickly filled once LAr
is filled. The LAr supply dewar is connected to the supply tank with a 
flex hose insulated with Cryo-Lite and the system dewar is slowly cooled
by flowing LAr into the bottom of the dewar. The flow is manually
regulated during the cool down to maintain the system pressure below the
8 psig threshold of the relief valve. It   takes about 1.5 hours to
begin accumulating LAr in the dewar.  It then requires another 20-30
minutes to fill the LAr to a level just below the lowest baffle plate.
Once the LAr dewar is filled, the condenser is filled with LN$_2$.
\item {\bf Continuous Operation} \\ 
The evaporation of the pressurized LN$_2$ in the condenser provides the
cooling power to maintain the system at the targeted LAr temperature.
Ideally, continuous and slow filling of LN$_2$ is desired to provide
constant cooling power to maintain a stable thermal condition in the
system. However, since the LN$_2$ transfer hose is not sufficiently well
insulated, this mode consumes much more LN$_2$ than an intermittent
filling cycle, in which the transfer hose is on average at a higher
temperature.  We have therefore made a compromise between the
desirability of thermal stability and the cost and nuisance of frequent
LN$_2$ dewar changes by operating in a intermittent fill mode.  In the
future we will obtain a vacuum jacked LN$_2$ transfer hose. A cryogenic
solenoid valve installed at the top of the condenser controls the
filling cycle. A batch fill is initiated by opening the solenoid valve
when the level of LN$_2$ in the condenser falls to a low limit, and is
terminated when the level of LN$_2$ reaches a high limit. The entire
batch filling process is controlled by the LabVIEW VI which compares the
value measured by the capacitance level gauge installed in the condenser
with stored limits, and opens and closes the solenoid valve accordingly.

Fig.~\ref{fig:lnlevel} shows a history of the LN$_2$ level, pressure,
and temperature during one day of operation. The low and high limits of
LN$_2$ level 
are set at 45\% and 80\% of the active length of the capacitance level
gauge. The back pressure regulator is set at 28.5 psig to maintain the
LN$_2$ temperature at 87.9 K. Each peak on the pressure curve marks the
beginning of a single fill. The increase in pressure during the fill is
caused by the large flow rate of gas introduced into the condenser by
the boiling LN$_2$ in the hose and the warm top of the condenser.  The
back pressure regulator that controls this pressure has a non-zero
compliance (pressure change per flow rate change), so with this large
flow increase the pressure increases.  The undershoot of the level below
the low limit is due to the delay between opening the solenoid valve and the
start of liquid arrival through the warm tubing into the condenser. The
following sharp rise in the level is the actual filling of LN$_2$.  The
slow decrease of the level is due to the evaporation of LN$_2$ caused by
condensation of the GAr as well as thermal leakage through the Cryo-Lite
cryogenic insulation. This evaporation rate of LN$_2$ is proportional to the
total heat load of the system, which is measured to be about 46 W.
Details of this measurement are discussed in Sec.~\ref{sec:heatload}. 

Fig.~\ref{fig:lnlevel} also indicates the temperatures of the LN$_2$ and LAr during 
one day of operation. During the filling, the LN$_2$ temperature
slightly increases as a result of the increase of its pressure. The LAr
temperature varies during the non-fill portion of the cycle due to the
increased temperature of the LN$_2$ during the fill followed by the slow
decrease of thermal conductance of the condenser as the level of LN$_2$
falls. The measurement time of electron transport properties in LAr is
typically $\sim$ 15 minutes.  During this time the temperature of LAr is
maintained stable within $\sim$ 0.1 K as shown in
Fig.~\ref{fig:lnlevel}, With the batch fill set between 45\% to 80\% of
the active length of the capacitance level gauge, a complete  filling
cycle is $\sim$ 1.5 hours as shown in Fig.~\ref{fig:lnlevel}.

\item {\bf System Purification} \\
Impurities inside the main dewar are continuously removed by passing GAr
through the purifier and returning the pure condensed LAr to the dewar.
As the molecular sieve and the copper tablets absorb impurities, they
become saturated and they rapidly cease to remove impurities.  When the
water and oxygen concentration  begin to increase during circulation,
the molecular sieve and the copper tablets must be regenerated to
restore their effectiveness.  To begin the regeneration process, the
purifier cylinder is heated (with commercial electrical heating tape) to
$\sim$ 190 $^{\circ}\rm C$. High purity argon gas is then flowed through
the purifier for several hours, until the water concentration in the
exhaust gas measured the Servomex DF-760e moisture analyzer falls below
100 part-per-million (ppm).  This removes most of the water from the
molecular sieve. Finally, the gas is changed to a mixture of high purity
argon gas with 2\% hydrogen, and this gas is flowed at a rate of 5
standard liters per minute until the water concentration of the exhaust
gas is reduced to a few ppm.  The complete process generally takes
between 24 to 48 hours.  \end{itemize}

\subsection{System Heat Load}\label{sec:heatload}
\begin{figure}[htbp] 
\centering
\includegraphics[width=0.5\textwidth, angle=0]{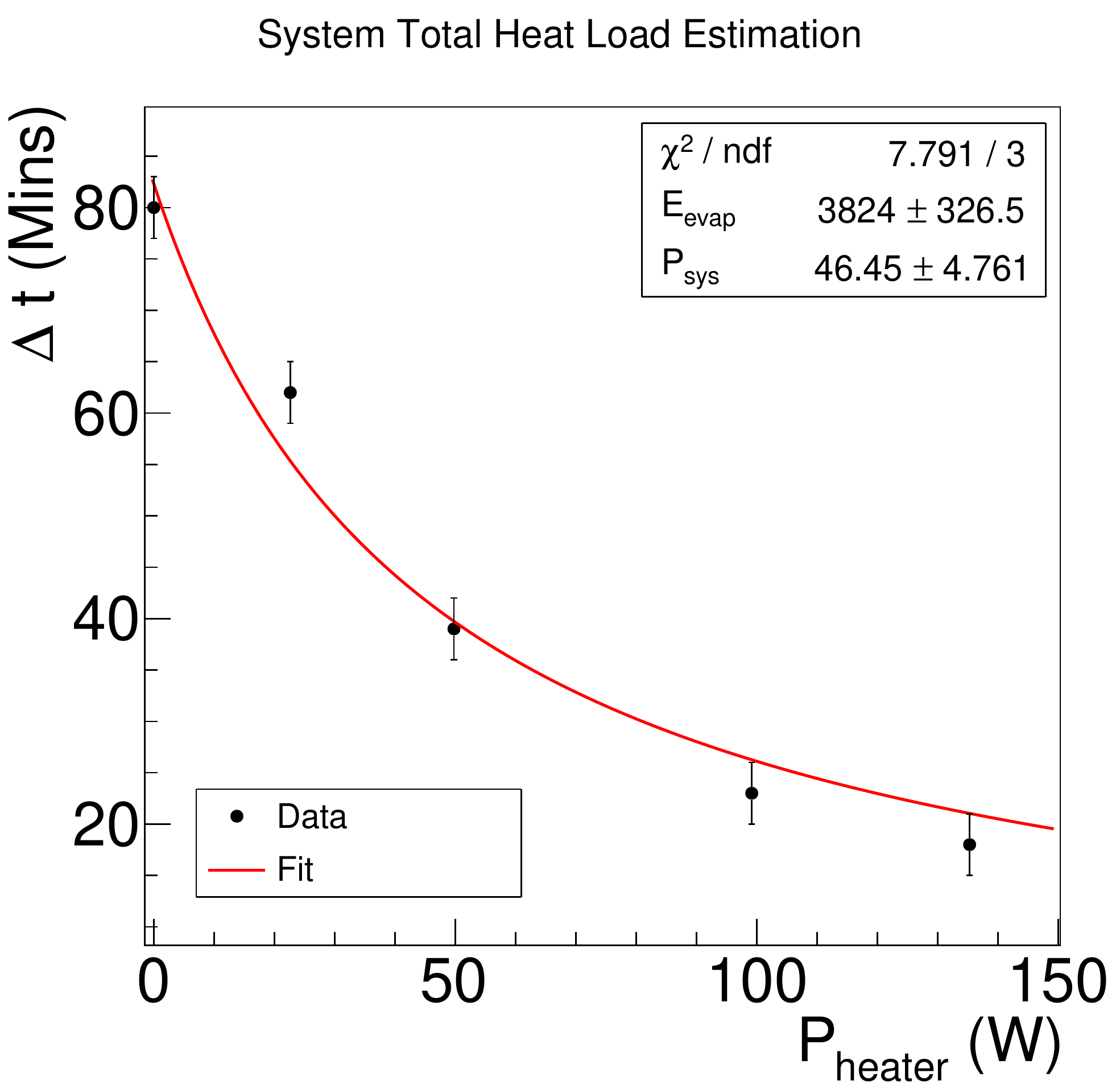}
\caption{Time between filling cycles is plotted against the amount
    of heat introduced to the system by the heater. Data are fitted with
    the function Eq. (3.1), 
        and the system's heat load is determined to be 46 $\pm$ 5 Watts.}
\label{fig:heatload}
\end{figure}

An understanding of the system's heat load is useful for optimizing
cryogenic operations.  The head load has been calculated with the Finite
Element Analysis (FEA) tool CFDesign using a model including materials,
insulation type, environmental boundary conditions, and GAr convection
in the gas volume. This calculation gives a total power input of 50
Watts.  This value is in agreement with two direct measurements. In the
first, the heat input is deduced from the rate of LN$_2$ consumed during
the evaporation portion of one filling cycle.  With the filling range
set at 45\% to 80\% of the active length of the capacitance level gauge,
1.6 L of LN$_2$ is evaporated during each cycle.  This is equivalent to
48 Watts at the latent heat of vaporization of LN$_2$ of $185.69$ J/g at
87 K. In the second measurement, the heat load is deduced from the
change in the duration of the evaporation cycle when additional heat
influx into the system. 
For this purpose, a heater immersed in the LAr (as described in
Sec.~\ref{sec:cryo_des}) is used.  The principle is as follows. During
each evaporation cycle, the total heat removed is $E_{evap} =
(P_{sys}+P_{heater}) \cdot \Delta t$, where $P_{heater}$ is the heat
introduced by the heater, $P_{sys}$ is the total heat load of the system
(with the heater off), and $\Delta t$ is the time between the opening
and closing of the
    solenoid valve. We thus expect
\begin{equation}\label{eq:formula}
P_{sys} = \frac{E_{evap}}{\Delta t} - P_{heater}.
\end{equation}
Therefore, the system's heat load $P_{sys}$ can be measured by varying
the heater power and measuring the period of a filling cycle. The result
of this measurement is shown in Fig.~\ref{fig:heatload}. The heat load
is found by this method to be 46 $\pm$ 5 Watts, consistent with the heat
load of 48 Watts obtained by the previous method.

\subsection{Argon Purity}

\begin{figure}[htbp]
\centering
\begin{minipage}[c]{0.53\linewidth}
\includegraphics[width=1.0\textwidth, angle=0]{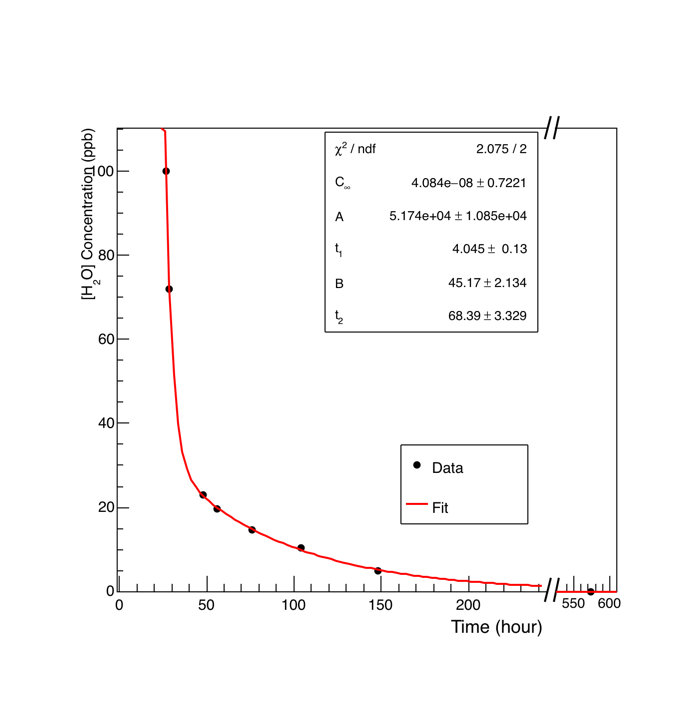}
\end{minipage}
\begin{minipage}[c]{0.45\linewidth}
\includegraphics[width=1.0\textwidth, angle=0]{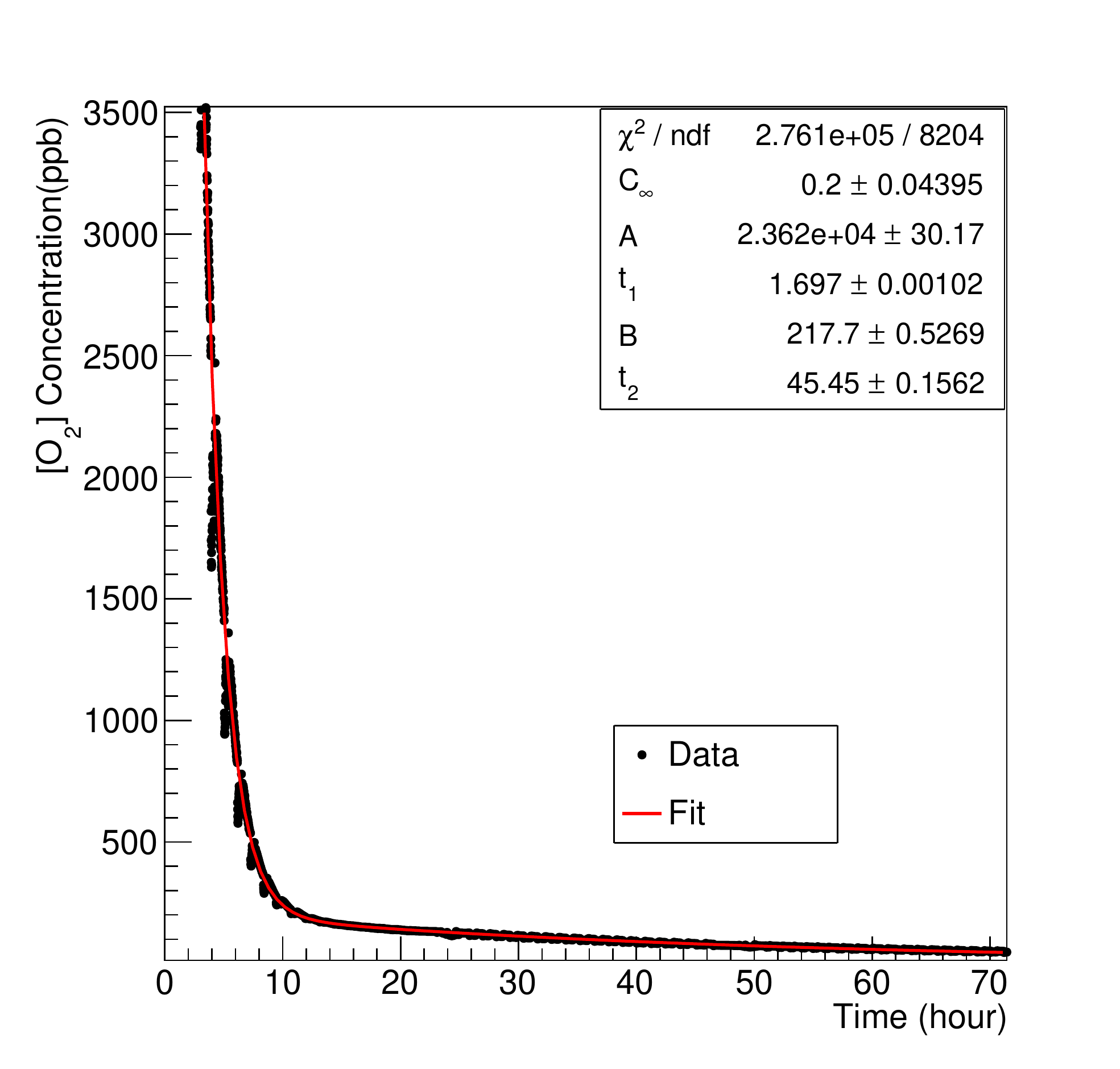}
\end{minipage}
\caption{Water (left) and oxygen (right) reduction curves of the system after 
the initial filling.}
\label{fig:cleaning}
\end{figure}

The mass exchange rate for gas purification alone is typically small
compared to the rate that can be obtained with liquid purification,
since cryogenic pumps can easily achieve large liquid flow rates through
purifiers in large LAr systems (such as LAPD~\cite{Adamowski:2014daa}).
However, the rate of outgassing of impurities depends strongly  on the
temperature~\cite{Andrews:2009zza}. At LAr temperature, the outgassing
rate is negligible. Impurities in the system are then contributed only
by outgassing from the surfaces of the GAr volume, which are  at a
higher temperature. Therefore, gas purification effectively
removes impurities at their source.  Although gas purification takes
longer to clean dirty LAr, it can in principle maintain high purity
liquid, provided that care is taken in the design of the system to
minimize the direct transfer of impurities from the gas into the liquid
(as, for example, by condensing gas directly into the liquid).

The water concentration is measured as a function of time after the
initial fill from the supply tank as shown in the left panel of
Fig.~\ref{fig:cleaning}. There are two exponential decay processes
during the purification. The first process, with a small time constant,
occurs just after the initial fill. This is caused by the initial purging of 
the sampling tubing and the detector
chamber volume in the gas analyzer. The second process, with a much
longer time constant, is the purification of the liquid volume.
The data are fitted by the function:
\begin{equation}\label{eq:cleaning}
    C_{H_2O}=C_{\infty}+A\cdot e^{-\frac{t}{t_1}}+B\cdot
    e^{-\frac{t}{t_2}},
\end{equation}
where $C_{H_2O}$ is the water concentration in the unit of ppb; $C_{\infty}\geq0$ is the ultimate water 
concentration at infinite cleaning time; $t$ is the time; 
$A$ and $B$ are the amplitudes for the slow and fast process in units
of ppb;  
$t_1$ and $t_2$ are the time constants of the fast and slow cleaning 
process, respectively. The fit gives the $t_1=4.05\pm0.13$ hours and
$t_2=68.4\pm3.3$ hours. The fact that the ultimate water
concentration approaches the limit of detection is an indication of the effectiveness 
of the gas purification system. Similarly, the measured oxygen
concentration as a function of time is shown in the right
panel of Fig.~\ref{fig:cleaning}.  The results demonstrate that gas
purification is sufficient for reducing the water and oxygen
concentration to $<$ 1 ppb. 
The achieved high purity LAr has also been confirmed in the
electron transport properties measurement, as the free electron lifetime
is found to exceed 400 $\mu$s. 

\begin{figure}[htbp]
\centering
\includegraphics[width=0.5\textwidth, angle=0]{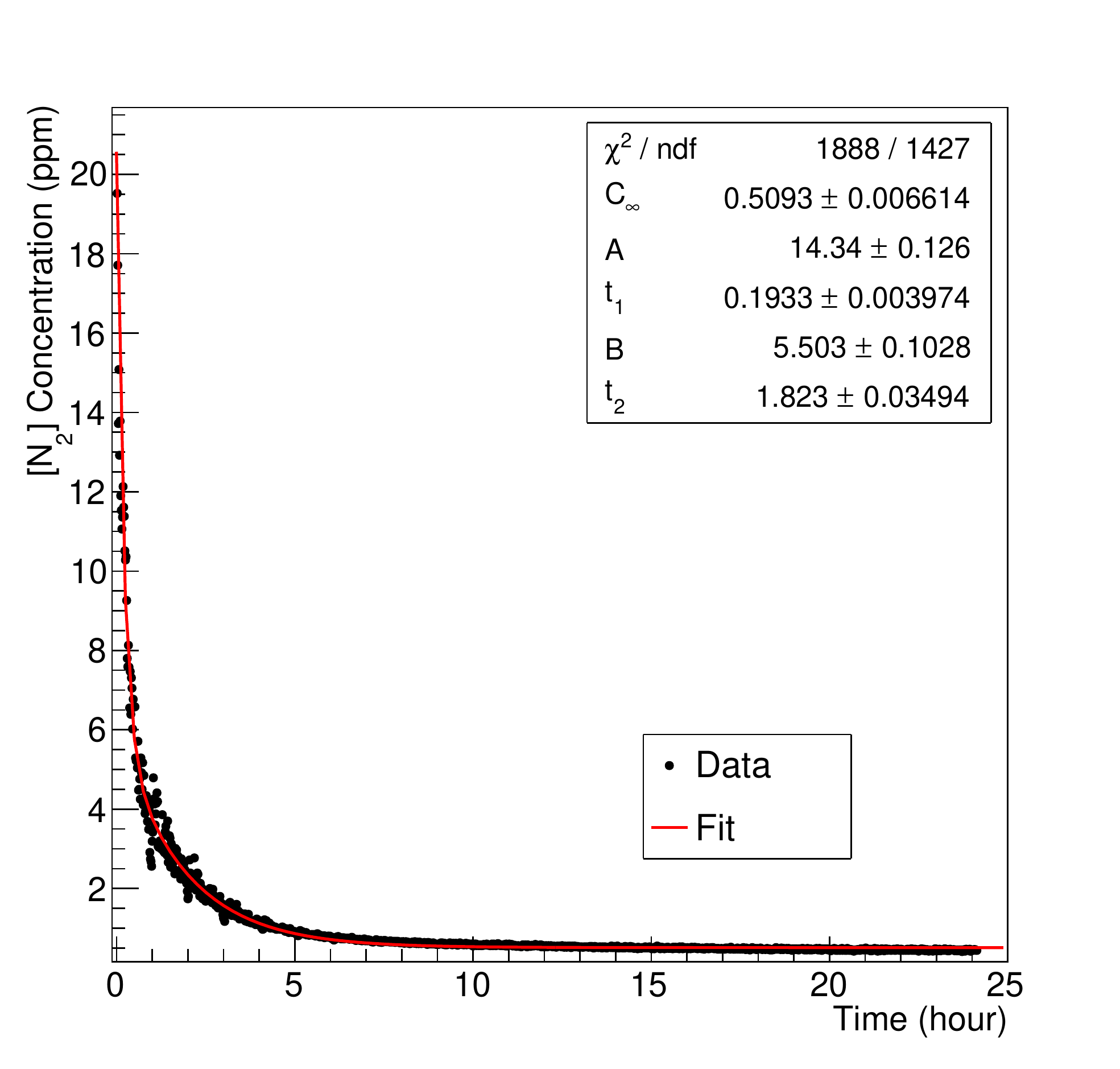}
\caption{Nitrogen reduction curve of system after the initial LAr filling.}
\label{fig:n2cleaning}
\end{figure}

The reduction of nitrogen concentration is shown in 
Fig.~\ref{fig:n2cleaning}. Although nitrogen has negligible influence on
the free electron lifetime, its concentration is limited in practical LArTPCs
to the order of a few ppm level in order to avoid quenching the production of 
scintillation light~\cite{Jones2013bca}.
A similar fit with Eq.~\eqref{eq:cleaning} shows 
$t_1=0.2$ hours and $t_2=1.8$ hours. The ultimate N$_2$ 
concentration approaches  0.5 ppm, which is several 
orders of magnitude higher than those of water and oxygen. 
This is due to the fact that the molecular sieve adsorbs much less nitrogen than water, and quickly saturates. 
Therefore, the ultimate nitrogen concentration  
depends primarily on the quality of the commercial LAr supply.

Typically, the commercial LAr contains 
impurity concentrations on the order of ppm for water, oxygen and
nitrogen. During operation, it typically takes about a week to reach the
required purity level of $<$ 1.0 ppb level for water and oxygen.  These
concentrations are further reduced with continuous gas purification and
ultimately reach the lowest detection levels of the analyzers.  The
effectiveness of the gas purification can potentially be used to
optimize the purification procedure in large LArTPCs. For example, gas
purification alone might be used as the primary method to maintain LAr
purity after initial liquid purification during filling, significantly
reducing the average power needed for the cyrogenic facilities. We note that
this is in contrast to present practise and
experience~\cite{Antonello:2014eha}. We intend to explore the limits of
this optimization with the construction of a 1-ton LAr system in the future.

\subsection{Photocathode Quantum Efficiency}
\begin{figure}[htbp]
\centering
\includegraphics[width=0.5\textwidth]{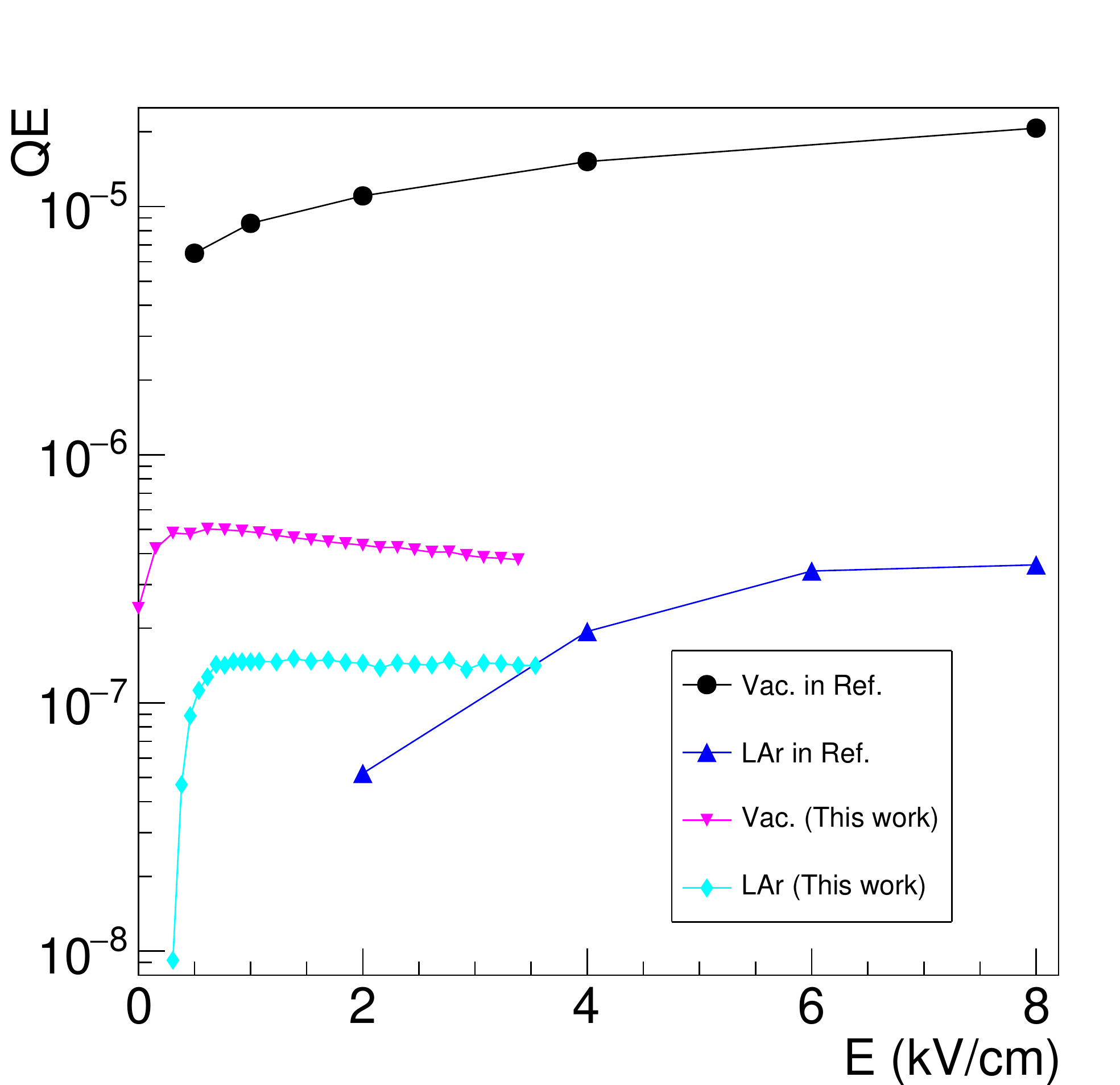}
\caption{The quantum efficiencies measured as a function of electric
    field at the
    minimum drift distance are compared with those 
of the same kind of photocathode in Ref.~\cite{Li:2015rqa}.
See text for more discussion. }
\label{fig:qe}
\end{figure}

The performance of the photocathode has been reported previously in
~\cite{Li:2015rqa} with a different experimental setup. The quantum
efficiencies (QE) measured for our gold photocathodes in  vacuum and LAr
conditions are shown in Fig.~\ref{fig:qe}. The QE is defined as the
number of electrons collected by the anode per number of UV photons
irradiating the back surface of the photocathode. The number of photons
is determined by the laser power, as measured by the power meter, and
the repetition rate. For the QE measurements, the anode is placed at the
minimum drift distance of 6.5 mm away from the photocathode in order to
minimize the impact of the finite free electron lifetime. The number of
electrons is then determined by the amplitude of the charge-sensitive
pre-amplifier signal.  The pre-amplifier is calibrated with a known
voltage pulse coupled through a known capacitor.

As shown in Fig.~\ref{fig:qe}, QE measured in vacuum with this apparatus
is significantly lower than reported in Ref.~\cite{Li:2015rqa}.  The
results in Ref.~\cite{Li:2015rqa} were obtained under two significantly
different conditions:  the entire setup is baked  up to $\sim$ 90
$^{\circ}\rm C$ and pumped to $\sim10^{-8}$ Torr, and the electron
current measured is leaving the cathode. Since our main 20-liter dewar
is vacuum insulated, no significant baking procedure is practical. In
addition, the 20-liter dewar can only be pumped down to $\sim 10^{-5}$
Torr with an order of magnitude larger volume (20 L vs. 2
L).~\footnote{The surface area of this system is significantly larger,
which naturally causes more outgassing.} Without baking and in the
poorer vacuum, surface contamination of the photocathode causes the QE
in vacuum to be lower in this setup.
However, since the contamination added by the argon is presumably
similar in both systems, the QEs in LAr are closer for the two sets of measurements. 
 The decline of QE at higher fields in our setup is presumably due to
 the decreasing transparency of the grid (absent in the results from
 Ref.~\cite{Li:2015rqa}) as the drift field increases. As shown in
 Sec.~\ref{sec:hvs}, we have calculated the transparency for
electrons drifting through our grid; a similar effect should occur in
vacuum.  The general increase in
observed QE with field (after accounting for grid transparency) as shown
by the QEs measured in Ref.~\cite{Li:2015rqa} 
is caused by backscattering of electrons in the gold (for the vacuum case) and,
additionally, in the LAr. It can be noticed that the QE increases faster
    in LAr than in vacuum.    Overall, the number of electrons produced per
pulse is $\approx$ 10$^5$ in LAr which is sufficient for the electron transport
properties studies.

\section{Discussion and Conclusion}\label{sec:future}

In this paper, we describe the design and operational performance 
of a 20-liter LAr test stand with a gas purification system. 
Excellent thermal stability has 
been achieved, which is crucial 
for studies of electron transport properties in LAr.
In addition, this system also demonstrates that high purity LAr 
can be obtained with only a gas purification system, which is simple in 
construction and low in cost. The measurements of electron transport 
properties, impurity exchange rates between GAr and LAr, and outgassing rates at low temperatures 
are continuing with this apparatus and will be reported in future publications.

\acknowledgments
This works is supported by Laboratory Directed Research and Development (LDRD)
of Brookhaven National Laboratory and U.S. Department of Energy, Office of Science, Office of High Energy
Physics and Early Career Research program under contract number
DE-SC0012704.

\section*{Appendix}
\begin{figure}[htbp] 
\centering
\includegraphics[width=1\textwidth,
    angle=0]{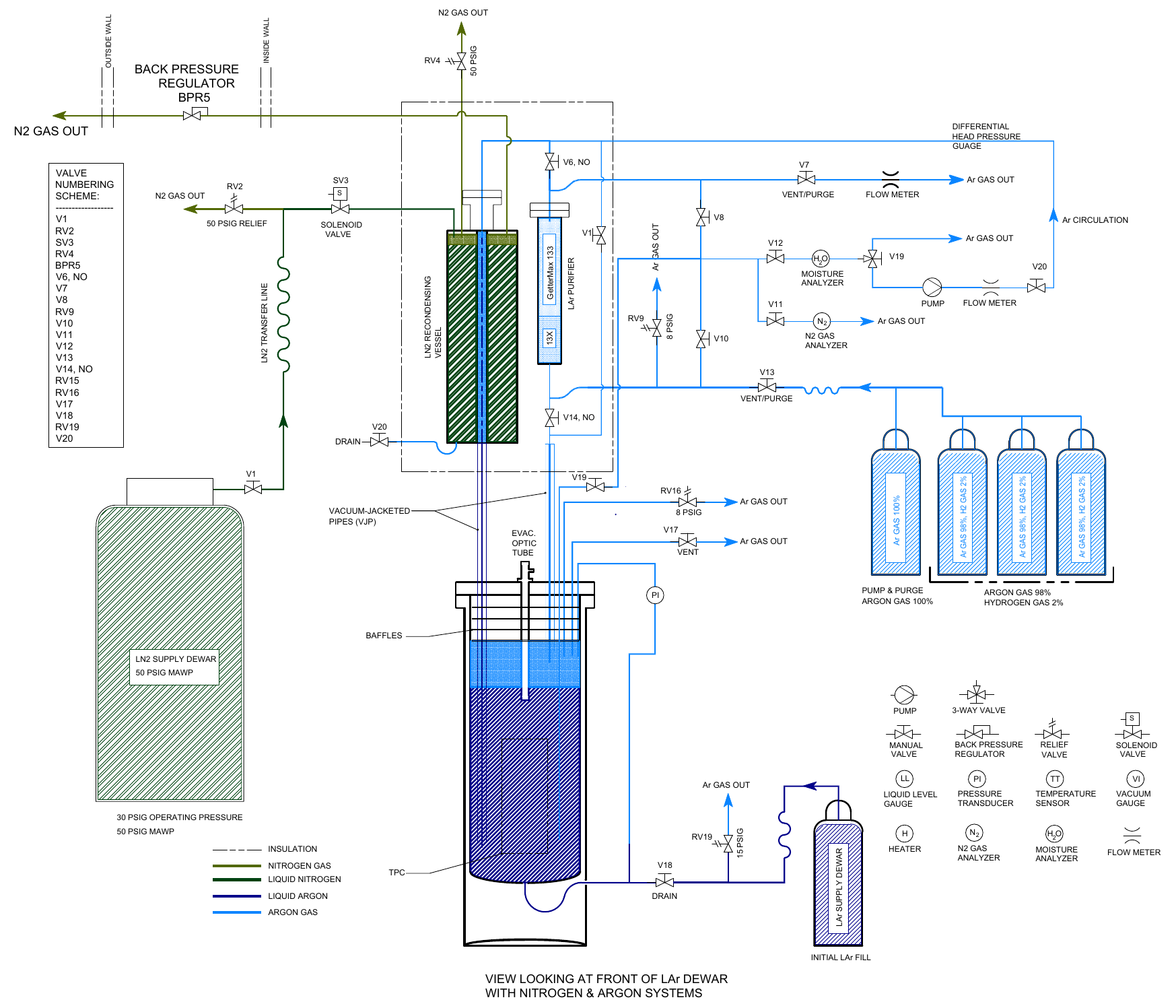}
\caption{P \& ID diagram of our setup.}
\label{fig:pid}
\end{figure}

The piping and instrumentation diagram (P\&ID) of our
system is shown in Fig.~\ref{fig:pid} for reference.

\bibliographystyle{hunsrt}
\bibliography{TS_JINST}{}

\end{document}